\documentclass[conference,12pt]{IEEEtran}
\usepackage{pslatex}
\usepackage{graphicx}
\usepackage{pst-node}
\newtheorem{theo}{\textbf{Theorem}}

\newtheorem{lemma}{\textbf{Lemma}}

\newcommand{\myN}{{\ensuremath{\mathrm{I\mkern-3.5mu N}}}}

\begin{document}

\title{A new queueing strategy for the\\ Adversarial Queueing Theory}

\author{
\authorblockN{Michael Hilker, Christoph Schommer}
\authorblockA{{\small University of Luxembourg, Campus Kirchberg}\\
{\small Dept. of Computer Science and Communication}\\
{\small 6, Rue Richard Coudenhove-Kalergi, L-1359 Luxembourg}\\
{\small Email: \{michael.hilker, christoph.schommer\}@ uni.lu}\\
{\small Phone: +352-420101-\{234,228\}}}
}


%


\maketitle

\thispagestyle{empty}

\begin{abstract}
In the today's Internet and TCP/IP-networks, the queueing of packets is commonly implemented using the protocol FIFO (First In First Out). Unfortunately, FIFO performs poorly in the Adversarial Queueing Theory. Other queueing strategies are researched in this model and better results are performed by alternative queueing strategies, e.g. LIS (Longest In System). 
This article introduces a new queueing protocol called \textit{interval-strategy} that is concerned with the reduction from dynamic to static routing. We discuss the maximum system time for a packet and estimate with up-to-date results how this can be achieved. 
We figure out the maximum amount of time where a packet can spend in the network (i.e. worst case system time), and argue that the universal instability of the presented interval-strategy can be reached through these results.
When a large group of queueing strategies is used for queueing, we prove that the interval-strategy will be universally unstable. Finally, we calculate the maximum time of the static routing to reach an universal stable and polynomial - in detail linear - bounded interval-strategy. Afterwards we close - in order to check this upper bound - with up-to-date results about the delivery times in static routing.

\end{abstract}
{\small
Keywords: (Adversarial) Queueing Theory, Internet Traffic Management, Data Streams.
} 

%
\IEEEpeerreviewmaketitle

\section{Introduction}
\label{secIntroduction}

In the today's Internet and TCP/IP-networks, the queueing of packets is commonly implemented using the protocol FIFO (First In First Out). However, when traffic resides heavily in the network, a queue overflow can occur and consequently a numerousness number of packets may become lost. Thereby, these packets must be resend by the transmitter; the outcome of this will be higher network traffic and higher delivery time for the resented packets as well as for all packets. The Quality of Life depends on fast, reliable and available communication connections. Without communication, e.g. internet, phone or email, is the today's life inconceivable. Queueing is one component which highly affects the speed and realibility of networks.

In this article\footnote{This work is supported by the \textit{Ministère Luxembourgeois de l'education et de la recherche} through the project \textit{INTRA (= INternet TRAffic management
and analysis)}, which is currently performed at the \textit{University of Luxembourg}. Parts of the work result from a cooperation with the \textit{Johann Wolfgang Goethe-University, Frankfurt/Main}.},
we concern with novel inspections on Queueing Theory in the Adversarial Queueing Model.
The \textit{Adversarial Queueing Model} was firstly investigated by Andrews et. al in 1996 \cite{And96} and
continued in 2001 \cite{And00a}. The Adversarial Queueing Model tries to simulate a real network in order to analyse queueing strategies theoretically. The Queueing is in the Adversarial Queueing Model simulated using a game between an adversarial - injects packets in order to overload connections - and the queueing strategy - tries to manage the packets injected by the adversarial. Queueing Theory in combination with deterministic adversaries
is a new and challenging research field. So far, the implementation of networks mostly uses the
FIFO protocol and seldom other alternatives.
 
In 2002, Bhattacharjee and Goel \cite{Bha03} published an article that described FIFO as an universal
unstable protocol, even at arbitrary low rates. Universal Unstable means that there exist an adversarial and a network where the number of packets in the system is not bounded. Generally, a lot of other basic
protocols like e.g. LIFO (Last In First Out), FFS (Farthest From Source), and NTG (Nearest To Go) are commonly known as to be universal unstable \cite{And00a}. Universal unstable protocols are inconvenient for the implementation of the queueing in networks because these protocols need infinite queue size as well as the delivery time of a packet is unbounded. 
On the other side, the protocols SIS (Shortest In System), FTG (Farthest To Go) and NTS (Nearest To Source) are universal stable, whereas the
upper bound for queue size is exponential for the longest path of the network \cite{And00a}. Even if a queueing strategy is universal stable it is also important that the number of packets in the system is polynomial bounded and not exponential; thus these queueing protocols are not convenient for queueing. The queueing protocol LIS (Longest In System) is 
universal stable, but a nontrivial upper bound - a bound must be in 
$o(maximum\mbox{ }pathlength\mbox{ }in\mbox{ }network)$ - is not currently known \cite{And00a, Adl02}. Furthermore, there exist a lot of alternative approaches for queueing in the Adversarial Queueing Model which are rarely analysed. In this paper, we inspect the novel and auspicious queueing protocol interval-strategy. The interval-strategy divides the packets injected by the adversarial in phases, transfers the packets of a phase and delays all packets which are injected during this phase until the next phase starts. 

The main goal in the Adversarial Queueing Model is firstly to find out whether a polynomial upper bounded protocol exists or not and secondly, if such a strategy can be used to describe how this polynomial upper bounded protocol behaves. The main motivation in the Adversarial Queueing Model is that the performance of existing connections is increased if other queueing strategies than the commonly implemented FIFO is applied. New applications based on internet are feasible with faster communication connections, e.g. HDTV over IP, increased network connection speed especially for home-connections, and/or high-bandwidth applications. 

We focus on the interval-strategy approach which reduces the dynamic routing to static routing by
reusing known facts about static routing in the context of Adversarial Queueing Theory. \\

The agenda of this article is:

Section \ref{secCurrentSituation} describes the Adversarial Queueing Model and some parameters to evaluate queueing protocols and section \ref{secDescribtion} defines the interval-strategy. In section \ref{secBasicNetworks}, we look on the behavior of the interval-strategy on basic networks, section \ref{secWorstCaseStatic} calculates the worst case time for static routing and section \ref{secNonForward} analyses the system time of static routing if non forward looking protocols - e.g. NTS, FFS - are used for queueing. In section \ref{secUpperBound}, we calculate the upper bound that static routing can use for a phase in order to get a universal stable and polynomial - in detail linear - bounded interval-strategy. Section \ref{secImprovment} improves the interval-strategy and section \ref{secConclusion} gives a conclusion and future prospect in this research area.

\section{Current situation}
\label{secCurrentSituation}
In the \textit{Adversarial Queueing Model} - introduced in \cite{And96} and \cite{And00a} - , the queueing is simulated by a game between the queueing protocol and an adversarial. The adversarial injects packets with the aim to increase the number of packets in the system. The queueing strategy manages which packet is allowed to traverse which connections in which time-step in order to limit the number of packets in the system. The Adversarial Queueing Model is used to model a realistic network situation such as it occurs in realized networks. The main benefit of this model is that a researcher can enforce a mathematic analysis without using probabilistic traffic. Hence, this model is nearer to practical network load; better results are expected. \\

Now take a look at the definition of the adversarial. The \textit{adversarial} has the right to inject packets with a fixed path through the network in every time-step. The number of packets injected by a $(r,b)$-adversarial in time interval $I$ for an edge $e$ is bounded by $r\cdot \vert I\vert+b$; $0<r<1$ is called \textit{injection-rate} and $b\ge 1$ is called \textit{burst}. The Adversarial Queueing Theory uses the parameter injection-rate in order to simulate the normal network-load as well as the parameter burst in order to simulate a short-term overload of a connection. Consequently, the adversarial is allowed to inject packets in the network over time and the queueing strategy has to guarantee that the packets arrive at the certain destination. The queueing strategy is evaluated how fast packets arrive at their destinations as well as how many packets are in a queue and in the system.

The adversarial knows the topology of the network as well as the behaviour of the queueing strategy. The aim of the queueing strategy is to guarantee a proper transmission of the packets as well as the aim of the adversarial is to increase the number of packets in the network. \\

To estimate, if a queueing protocol is convenient, we examine the parameter \textit{stability}. A protocol is \textit{stable} on a network against
an adversarial if the number of packets in the network is upper bounded in every time step. A queueing protocol is \textit{universal stable} if the protocol
is stable against all adversaries on all networks. Universal stability of a queueing strategy means that in all time steps is the number of packets in each queue upper bounded on all networks against all adversaries. If a queueing strategy is universal stable, it is also important to examine the number of packets in the system; only if this value is polynomial bounded - the global aim is to have a linear value - is the strategy appropriate for queueing. Universal instability of a queueing strategy means that there exist a network and an adversarial so that the number of packets in the system is in the $i$-th time step greater than in the $i-1$-th time step ($i\in\myN$). These universal unstable queueing strategies are inconvenient because the queueing strategies need queues with an infinite queue size.

There are two main ways of proving. If the motivation is 

\begin{itemize}
   \item to demonstrate that a queueing protocol is \textit{universal unstable}, we have to define a network and an adversarial such that the number of packets in the network is not upper bounded. 
   \item to demonstrate that a queueing protocol is \textit{universal stable}, we need to upper bound the number of packets for 
         all networks and adversaries. Usually it is also interesting to estimate the value of the bound. 
\end{itemize}

Unfortunately, the average case analysis is less discussed in the research community. The main reason
for this is a missing definition and the continuing focus on an existing idea, namely to find
a queueing strategy that probably satisfies all occurring problems through one network and adversarial.
Therefore, the main approaches in Adversarial Queueing Model base on worst case analyses.

\section{Description of interval-strategy}
\label{secDescribtion}
As described in the definition of the Adversarial Queueing Model, the adversarial injects packets over time. In order to receive a good delivery time of the packets, the packets are divided by the interval-strategy in data phases. Packets pertaining to a phase are routed and all packets of later phases are delayed. The routing of one phase is called \textit{static routing} because all packets reside in the network at startup. Therefore, the static routing represents a special situation of dynamic routing. \\

For \textit{dynamic routing}, a queueing strategy operates on a network and against an adversarial which injects packets over time. Hence, the information for the queueing strategy can change every time step. 

The packets traverse their paths through the network that is fixed by the adversarial.

In each queue and at every time step, the queueing strategy decides which packet traverses the edge of the queue. Accordingly, this is the normal situation in queueing and routing theory. \\

For \textit{static routing}, however, all packets reside in the network at startup and no packets will be injected after startup; all information is existing and applicable for the queueing strategy from beginning. Of course, static routing is easier to handle as dynamic routing as well as there exist approaches with better results for static routing. \\

The idea of the interval-strategy is to reduce the dynamic routing to static routing because static routing is easier to handle. To map the dynamic routing to static routing, all network edges own two queues, respectively:
\begin{enumerate}
\item The first queue is the queue that cares for the static routing.
\item The second queue contains the packets injected during a static routing phase until the static routing phase ends. \\
\end{enumerate}
Concerning the description of the interval-strategy, it will be assumed that there are no packets in the network at startup. Afterwards, the adversarial injects some packets in the network in one time-step. These packets are the packets which are routed by the first static routing phase. 

The interval-strategy transmits the packets of the first static routing phase to their destinations and all packets which are injected during the static routing phase are delayed in the second queue. If all packets of the first static routing phase reach the destinations and the first static routing phase ends, the packets delayed in the second queues are copied to the first queues and the second static routing starts with these packets.

Thus, if the $i$-th static routing phase is running, these packets will be routed and new injected packets will be delayed in the second queues. If all packets of the $i$-th path receive their destinations the packets of the second queues will be copied to the first queue and, afterwards, the $i+1$-th static routing phase is started with these packets. \\

At startup time of the network, we notice a static routing phase without any packets. In this case, it terminates immediately and the first phase starts routing the packets injected during the first time step.
So, the interval-strategy uses a static routing phase for delivering the packets.\\

With some modifications, we know that the interval-strategy performs well \cite{And00a,Sch04}. These modifications are:
\begin{itemize}
\item	the interval-strategy delays all injected packets for a probabilistic time period and
\item	if more than one packets want to traverse a edge the decision which packet is allowed to traverse the edge is performed by a probabilistic event.
\end{itemize}
This probabilistic version of the interval-strategy is with high probability universal stable as well as the number of packets in system is with high probability polynomial bounded.

The idea of the interval-strategy is to use similar requirements and to receive similar results towards a deterministic strategy. However, the interval-strategy is a deterministic version of the probabilistic. Fortunately, there exist some indicators that the interval-strategy is almost reliable since the similar random-based strategy offers a well performance. We believe that there exist a schedule for static routing so that the interval-strategy is universal stable and polynomial - in detail linear - bounded. Furthermore, a queueing strategy that deterministically performs the mapping from dynamic to static routing was rarely analysed in the Adversarial Queueing Theory. \\

Furthermore, we implemented the interval-strategy on a network simulator and the results are promising because the number of packets in the system is always bounded in a lot of scenarios. To simulate a scenario, a network and adversarial have been designed; they have produced lots of packets in any queues. We notice that the results of the simulations substantiate the theoretical results of this article.
 
\section{Interval-strategy on basic networks}
\label{secBasicNetworks}
In this section we describe the behavior of the interval-strategy on basic networks.

\subsection{One way connection line}
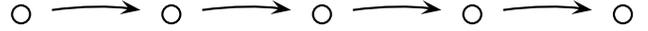
\begin{figure}[t]
\begin{center}
\begin{pspicture}(1,1)(9,3)
\cnodeput(1,1){1}{}\cnodeput(3,1){2}{}\cnodeput(5,1){3}{}\cnodeput(7,1){4}{}\cnodeput(9,1){5}{}
\ncarc[nodesep=8pt, arrowsize=3pt 3]{->}{1}{2}
\ncarc[nodesep=8pt, arrowsize=3pt 3]{->}{2}{3}
\ncarc[nodesep=8pt, arrowsize=3pt 3]{->}{3}{4}
\ncarc[nodesep=8pt, arrowsize=3pt 3]{->}{4}{5}
\end{pspicture}
\end{center}
\caption{one way connection line} \label{FigOneWayConLine}
\end{figure}

One way connection lines are basic networks where some nodes can send packets from A to C via B, but C and B cannot send packets to A. If A is the start of the line, B is the end of the line and we insert an edge $e=(B,A)$ we will receive a ring; an example network is visualized in figure \ref{FigOneWayConLine}. In this part we look on the performance of the interval-strategy on one way connection lines.

Let $A=(b,r)$ the adversarial and let $G$ a one way connection line with length $d$. 

\begin{enumerate}
\item In the first static routing phase is the number of packets upper bounded by $r\cdot 1+b$ which is equal to $b$ because $r\cdot 1<1$ and a packet cannot be fragmented. Hence we see that the static routing phase needs up to $b+d$ time steps because $b$ packets can interfere and every packet needs up to $d$ time steps to traverse the own path.
\item In the second phase the packets - injected during the first phase - will be routed. The number of these packets is upper bounded by $r\cdot b+r\cdot d$ and the second phase needs up to $r\cdot b+r\cdot d+d$ time steps.
\item[i)]  We see iteratively that the number of packets in the $i$-th phase is upper bounded by 
$$r^{i-1}\cdot b+\sum\limits_{j=1}^{i-1}{r^j\cdot d}$$
Hence the number of time steps needed by the $i$-th phase is upper bounded by
$$r^{i-1}\cdot b+\sum\limits_{j=0}^{i-1}{r^j\cdot d}$$
\end{enumerate}

If we look on $i\rightarrow\infty$ we receive a time for static routing of 
$$\frac{d}{1-r}$$

Therefore, the maximum delivery time of a packet is bounded by
$$2\cdot\frac{d}{1-r}$$
because in the worst case, the packet waits maximum time in the second queue and afterwards the packets needs the maximum time for routing.

Hence, the interval-strategy is universal stable and polynomial bounded - in detail linear - if the network is a one way connection line.
\subsection{Trees}
\begin{figure}[t]
\begin{center}
\begin{pspicture}(0,1)(7,7)
\cnodeput(0,1){o0}{}
\cnodeput(1,1){o1}{}
\cnodeput(2,1){o4}{}
\cnodeput(3,1){o2}{}
\cnodeput(6,1){o3}{}
\cnodeput(2,3){m1}{}
\cnodeput(5,3){m2}{}
\cnodeput(6.5,3){m3}{}
\cnodeput(4,5){M}{}
\cnodeput(4,7){O}{}

\ncarc[nodesep=8pt, arrowsize=3pt 3]{->}{o0}{m1}
\ncarc[nodesep=8pt, arrowsize=3pt 3]{->}{o1}{m1}
\ncarc[nodesep=8pt, arrowsize=3pt 3]{->}{o4}{m1}
\ncarc[nodesep=8pt, arrowsize=3pt 3]{->}{o2}{m1}
\ncarc[nodesep=8pt, arrowsize=3pt 3]{->}{o3}{m2}
\ncarc[nodesep=8pt, arrowsize=3pt 3]{->}{m1}{M}
\ncarc[nodesep=8pt, arrowsize=3pt 3]{->}{m2}{M}
\ncarc[nodesep=8pt, arrowsize=3pt 3]{<-}{M}{m3}
\ncarc[nodesep=8pt, arrowsize=3pt 3]{->}{M}{O}
\end{pspicture}
\end{center}
\caption{example for a tree} \label{FigTree}
\end{figure}
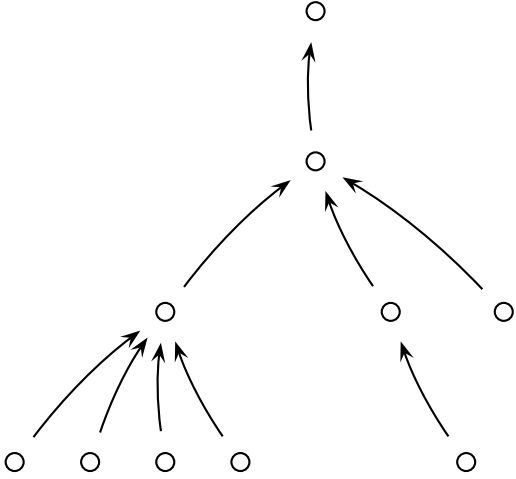

Here we focus our view on the performance of the interval-strategy on trees if static routing needs the worst case time. An example for a tree is visualized in figure \ref{FigTree}.

Let $A=(b,r)$ an adversarial with parameters $b\ge 1$ and $0<r<1$. Let $G$ a network with $d$ as the length of the longest path in the network. And let $n\cdot d$ the system time of static routing with $n$ the maximum number of packets which need an edge $e$ (worst-case system time for static routing, similar to section \ref{secWorstCaseStatic}).

\begin{enumerate}
\item In the first phase the number of packets is upper bounded by $b$ - definition of adversarial $A$ - and hence the system time of the static routing is upper bounded by $b\cdot d$. 

\item The second static routing phase contains packets which are inserted in the first phase. Therefore, the number of packets in the second phase is upper bounded by $r\cdot b\cdot d$. Consequently, the system time of the second phase is upper bounded by $r\cdot b\cdot d^2$.

\item[i)] We can see iteratively that the number of packets in the $i$-th phase is upper bounded by $r^{i-1}\cdot b\cdot d^{i-1}$ and hence the static routing phase is upper bounded by $ r^{i-1}\cdot b\cdot d^i$. 
\end{enumerate}

Now we look on $i\rightarrow\infty$:
\begin{eqnarray*}
\lim\limits_{i\rightarrow\infty}{\left(r^{i-1}\cdot d^i\cdot
b\right)} & \left\{
\begin{array}{*{1}{l}}
= 0\\
= d\cdot b\\
\rightarrow\infty
\end{array}
\begin{array}{*{1}{c}}
\mbox{, if } r\cdot d<1\\
\mbox{, if } r\cdot d=1\\
\mbox{, if } r\cdot d>1
\end{array}
\right.
\end{eqnarray*}

We see that the interval-strategy is universal unstable on trees if the practical case $r\cdot d>1$ appear. In the other cases, $r\cdot d\le 1$, is the longest network length short and the adversarial performance is low; hence, these cases are not near to practice. Afterwards, we calculate the worst case of static routing, thereafter, we analyse a common used group of protocols for queueing and, accordingly, we calculate the maximum time static routing may use to receive a universal stable and polynomial bounded interval-strategy.
 
\section{Worst case system time \\ in static routing}
\label{secWorstCaseStatic}
In this part, the worst-case system time of a packet in static routing is exhibited. In static routing, all packets reside initially in the network. A static routing phase ends when all packets reach their final destinations. \\

\begin{lemma}
\label{lemma1}
Let $n$ be the maximum number of packets for an edge in the network and let $d$ be the length of the longest path in the network. Then the system time is bounded by $n\cdot d$.
\end{lemma}

\textbf{Proof:} If a packet needs $n\cdot d+1$ time steps to reach the destination, it will either be blocked in the queue of any edge by $n+1$ packets or it will be blocked in every queue by $n$ packets and traverses $d+1$ edges. The former case is a contradiction to the definition of the parameter $n$, the latter case a contradiction to the definition of the parameter $d$. Hence, the lemma \ref{lemma1} is proved.\\

With this lemma \ref{lemma1}, we notice that a static routing phase ends not later than $n\cdot d$ time steps. This is the worst case time for static routing.\\

\begin{lemma}
\label{lemma2}
The interval-strategy will be universal unstable if the worst case is applied for static routing.
\end{lemma}

\textbf{Proof:} The first static routing phase contains up to $b$ packets. So, the static routing phase needs $b\cdot d$ time steps. In these time steps, the packets for the second phase are injected. The packets in the second phase are upper bounded by $r\cdot b\cdot d$; the static routing phase consequently needs $r\cdot b\cdot d^2$ time steps. If this will be iterated, the number of packets in the $i$-th phase is upper bounded by $r^{i-1}\cdot b\cdot d^{i-1}$ and the system time for the $i$-th phase is upper bounded by $r^{i-1}\cdot b\cdot d^i$. If $i\rightarrow\infty$ and $r\cdot d>1$ is valid, the number of packets in the interval-strategy will be unbounded and the interval-strategy will be universal unstable. The special case of $r\cdot d\le 1$ remains uninteresting because the injection rate is rather small and the longest path in the network is short.

\textbf{Remark:} It is easy to design a network and an adversarial where the worst case time for static routing occurs in every phase.\\

With both lemma \ref{lemma1} and \ref{lemma2}, the worst case time for static routing is too slow for the interval-strategy in order to become convenient. Therefore, we need to find better solutions for static routing in order to evaluate these solutions for the interval-strategy. 

\section{Interval-strategy with \\non forward looking protocols}
\label{secNonForward}
\textit{Non forward looking protocols} are queueing strategies which only evaluate the bygone paths that were visited by the packets. They do not evaluate the given paths which the packets will traverse in the future. Examples for such protocols are NTS and FFS; with a few modifications, the proof is also valid for alternatives like LIS and SIS.

In the year of 1994, Leighton, Maggs and Richa showed that there exist a network and an adversarial where non forward looking protocols have a system time of at least $\frac{n\cdot (d-1)}{log(n)}$ for static routing \cite{Lei94}. This result will be used as a validation that these protocols do not satisfy the interval-strategy.\\
Here, a tree is used with no preconditions. Each static routing phase is disjoint and a system time greater or equal $\frac{n\cdot (d-1)}{log(n)}$ appears in every phase of the interval-strategy.
 
We will show iteratively that the interval-strategy is universal unstable and that these protocols are inconvenient for the protocol. Therefore, we assume that all static routing phases will have the system time of $\frac{n\cdot (d-1)}{log(n)}$.

In the first phase, there are only packets in the network which are injected during the first time step; thus, the number of these packets is bounded by $b$ and, consequently, the system time is $\frac{b\cdot (d-1)}{log(b)}$.

All packets which are routed in the second phase are injected during the first phase; hence, the number of packets is bounded by $r\cdot\frac{b\cdot (d-1)}{log(b)}$ and the system time for the static routing phase is 
$$\frac{r\cdot\frac{b\cdot (d-1)}{log(b)}\cdot (d-1)}{log(r\cdot\frac{b\cdot (d-1)}{log(b)})}$$

Let 
$$k_1=\frac{b\cdot(d-1)}{\log(b)}$$ 
and 
$$k_i=\frac{r^{i-1}\cdot(d-1)^i\cdot b}{\log(b)\cdot\prod\limits_{j=1}^{i-1}{\log(k_j)}}$$
then we see that $k_i$ is the system time for static routing in the $i$-th phase.

If $i\rightarrow\infty$ and $r\cdot(d-1)>1$, both the number of packets and the system time of the static routing phase will be unbounded. The special case of $r\cdot(d-1)\le 1$ is far away from practice because the injection rate $r$ is rather small and the longest network path is short.

From there, these non forward looking protocols are not suitable for implementing the queueing in static routing under practical conditions. In order to receive better results, the maximum system time for static routing will be calculated. Up-to-date result of static routing qualifies the interval-strategy.

\section{Upper bound for static routing for a polynomial bounded interval-strategy}
\label{secUpperBound}
In this section, a sharp upper bound for the static routing phase system time is calculated in order to receive an universal stable and polynomial bounded interval-strategy. We show as well that if the interval-strategy is polynomial bounded it is also linear bounded. It follows:
\begin{itemize}
		\item If this upper bound is reached by a protocol in static routing, the interval-strategy is universal stable and polynomial bounded. 
		\item If this upper bound is not accomplishable, the interval-strategy is not polynomial bounded and perhaps universal unstable. For this, the interval-strategy is no longer convenient.\\
\end{itemize}

\begin{theo}
If the static routing can be upper bounded by $c_1\cdot n+c_2\cdot d+c_3$, the interval-strategy will be universal stable and polynomial bounded. The parameters are defined as follows:
	\begin{itemize}
		\item	$c_1$, $c_2$ and $c_3$ are parameters with a constant value, $c_1\le 1$,
		\item	$n$ is the maximal number of packets which contain an edge $e$ in the path and
		\item	$d$ is the length of the longest path in the network.
	\end{itemize}
\end{theo}

\textbf{Proof:}
By induction with three different static routing phases to verify the base cases, the predication will be proved.

\begin{enumerate}

\item Startup and first phase:

No packets are located in the system at startup. During one time step, an adversarial can inject at most $r\cdot 1+b$ packets for one edge; therefore, $n$ is at most $b$ because $r\cdot1<1$. The duration of the first static routing phase is upper bounded by $c_1\cdot b+c_2\cdot d+c_3$.

\item Second phase:

All packets which will be routed in the second phase are injected during the first phase. Thereby, the number of packets for one edge is upper bounded by $r\cdot(c_1\cdot b+c_2\cdot d+c_3)$ and hence, the static routing phase is upper bounded by 
$$c_1\cdot r\cdot(c_1\cdot b+c_2\cdot d+c_3)+c_2\cdot d+c_3=$$
$$r\cdot c_1^2\cdot b+r\cdot c_1\cdot c_2\cdot d+r\cdot c_1\cdot c_3+c_2 \cdot d+c_3$$

\item Third phase:

We observe the following: all routed packets of the third phase were inserted in the second phase. The number of these packets is bounded by 
$$r\cdot (r\cdot c_1^2\cdot b+r\cdot c_1\cdot c_2\cdot d+r\cdot c_1\cdot c_3+c_2\cdot d+c_3)$$
and the system time for the third static routing phase is bounded by
$$r^2\cdot c_1^3\cdot b+r^2\cdot c_1^2\cdot c_2\cdot d+r^2\cdot c_1^2\cdot c_3+$$
$$r\cdot c_1\cdot c_2\cdot d+r\cdot c_1\cdot c_3+c_2\cdot d+c_3$$
\end{enumerate}

\begin{itemize}
\item[i)] i-th phase:

By induction we see that the number of packets in the $i$-th phase is bounded by
$$r^{i-1}\cdot c_1^{i-1}\cdot b+\sum\limits_{j=1}^{i-1}{\left[r^j\cdot c_1^{j-1}\cdot\left(c_2\cdot d+c_3\right)\right]}$$
and the system time of the $i$-th static routing phase is bounded by 
$$r^{i-1}\cdot c_1^i\cdot b+\left(c_2\cdot d+c_3\right)\cdot\frac{(r\cdot c_1)^i-1}{r\cdot c_1-1}$$

\end{itemize}

If $i\rightarrow\infty$, the interval-strategy will be polynomial because out of $r\cdot c_1\le 1$ follows $(r\cdot c_1)^i\le 1$ as well as $c_2$ and $c_3$ are constant. If $r\cdot c_1>1$ or either $c_2$ or $c_3$ is not constant, the queue size of the interval-strategy will be at least exponential. The Theorem is shown.
 
The question is now is the interval-strategy also linear bounded - in the length of the longest path $d$? Thus, we have 
$\lim_{i\rightarrow\infty}r^{i-1}\cdot c_1^i\cdot b+\left(c_2\cdot d+c_3\right)\cdot\frac{(r\cdot c_1)^i-1}{r\cdot c_1-1} = \Theta(d)$
and it follows that $0\le r\cdot c_1\le 1$ and $c_2\cdot d + c_3 = \Theta(d)$ must be valid. Are these equations valid? The former equation is valid because the values of the parameters $c_1$ and $r$ are in $[0,1]$. The latter equation is valid because $c_2$ and $c_3$ are constant. Hence, if the interval-strategy is polynomial bounded the strategy is also linear bounded.

We have to check now if this system time for the static routing can be achieved. Therefore, we present up-to-date facts of static routing. 

It is highly important that in \cite{Lei94} is proved that a schedule for static routing with time $O(n+d)$ exists for every packet set. Furthermore, there are no networks and adversaries known with a time greater than $n+d$. Unfortunately, this upper bound is not proved. Since the proof of \cite{Lei94} is not constructive, it is open in which time the schedule with time $O(n+d)$ can be computed. There exist some promising algorithms that calculate a schedule of acceptable system time, but these algorithms are probabilistic that can not be used in this deterministic model. Indeed, there are no deterministic protocols for static routing with acceptable system time known.

The analysis of the interval-strategy has finished, but we still have to search for acceptable and deterministic algorithms to solve static routing.

\section{Improvement of interval-strategy}
\label{secImprovment}
The following case can occur: A static routing phase is running. There are no packets in the static routing phase which need an edge $e$ and the adversarial injects some packets for this edge $e$. Now these packets will be delayed until the static routing phase finishes and, unfortunatly, the edge has no traffic load. Hence, these packets can traverse the edge $e$ without interfering with the packets of the static routing phase if the packets will inserted into the second queue at the destination of edge $e$. Therefore, the interval-strategy will be improved because in some cases the delivery time for a packet decreases.	

\section{Conclusion}
\label{secConclusion}
The problem of novel queueing strategies becomes crucial and of particular importance
because the common used protocol FIFO poorly performs in the Adversarial Queueing Model.
In this context, there might be some hope to increase the network performance by alternative
queueing protocols. The benefits of higher network performance without increasing the connection speed are a better QoS (Quality of Service), cheaper high speed connections and a better reliability for network and communication connections. The Quality of Life may increase if faster communication connections are available. The price of faster communications connections decreases and the connections are obtainable for anybody. 

In this article, a new approach for queueing has been introduced, described and, in respect to the protocol, analysed. As probably the most important result, the maximum system time for static routing is calculated and compared
with up-to-date results about static routing. In order to prove the appropriateness and the goodness of the presented
interval-strategy, additional studies concerning static routing need to be done.

There exist further interesting questions that concern e.g. the lower bounds for static routing
(which system time does a static routing phase need for a given network and adversarial) and/or
the important question of this model: does there exist a queueing strategy which offers a polynomial or even linear bounded
delivery time for a packet and - in the case it exists - what is it's behavior? Additionally, it would become worthful
to examine a network simulation under realistic conditions using an unconventional strategy for queueing.
These questions will be our challenge for future researches.





%


\bibliography{Paper}
\bibliographystyle{plain}
\end{document}